\documentclass[final,3p,twocolumn,10pt]{elsarticle}

\usepackage{graphics}
\usepackage{graphicx}

\journal{Physica E}

\begin{document}

\begin{frontmatter}



\title{Gate Adjustable Coherent Three and Four Level Mixing in a Vertical Quantum Dot Molecule}


\author[NRC,McGill]{C. Payette \corref{cor1}}
\cortext[cor1]{Corresponding author: Phone: +1-613-993-8133, FAX: +1-613-950-0202}
\ead{chris.payette@nrc-cnrc.gc.ca}
\author[ICORP]{S. Amaha}
\author[ICORP]{T. Hatano}
\author[RIKEN]{K. Ono}
\author[NRC]{J.A. Gupta}
\author[NRC]{G.C. Aers}
\author[NRC,McGill]{D.G. Austing}
\author[UofT]{S.V. Nair}
\author[ICORP,Tokyo]{S. Tarucha}
\address[NRC]{Institute for Microstructural Sciences, National Research Council of Canada, Ottawa, Ontario K1A 0R6, Canada}
\address[McGill]{Department of Physics, McGill University, Montreal, Quebec H3A 2T8, Canada}
\address[ICORP]{Quantum Spin Information Project, ICORP, JST, Atsugi, Kanagawa 243-0198, Japan}
\address[RIKEN]{RIKEN, Wako, Saitama 351-0198, Japan}
\address[UofT]{Center for Advanced Nanotechnology, University of Toronto, Toronto, Ontario M5S 3E3, Canada}
\address[Tokyo]{Department of Applied Physics, University of Tokyo, Tokyo 113-8656, Japan}

\begin{abstract}
We study level mixing in the single particle energy spectrum of one of the constituent quantum dots in a vertical double quantum dot by performing magneto-resonant-tunneling spectroscopy. The device used in this study differs from previous vertical double quantum dot devices in that the single side gate is now split into four separate gates. Because of the presence of natural perturbations caused by anharmonicity and anistrophy, applying different combinations of voltages to these gates allows us to alter the effective potential \emph{landscape} of the two dots and hence influence the level mixing. We present here preliminary results from one three level crossing and one four level crossings high up in the energy spectrum of one of the probed quantum dots, and demonstrate that we are able to change significantly the energy dispersions with magnetic field in the vicinity of the crossing regions. 
\end{abstract}

\end{frontmatter}
\section{Introduction}
\label{intro}
Means to realize strong mixing of three or more energy levels to investigate quantum coherent transport are required for advanced quantum information protocols in the solid state. For example, Refs. \cite{MichaelisEPL2006} and \cite{GreentreePRB2004} describe interesting all-electrical single-electron-tunneling schemes for coherent population trapping and coherent tunneling by adiabatic passage. Vertical quantum dots (QDs) possessing a high degree of symmetry \cite{TaruchaPRL1996,MatagnePRB2002} offer unexpected opportunities to observe strong mixing between multiple single-particle levels over a wide energy window \cite{OnoPhysB2002,PayettePRL2009}. Recently, we measured and successfully modeled three-level coherent mixing, attributed to natural anharmonicty and anisotropy in the QD confinement potentials, leading for instance to resonant current suppression (``dark'' state formation) in a vertical double QD device with a single side gate wrapped around the circular mesa \cite{PayettePRL2009}. However, in this earlier work, it was not possible to alter the effective QD confinement potentials in-situ and so the mixing could not be influenced experimentally. In this work, we now use multiple-gate technology \cite{AustingSemi1997} to build a vertical double QD device in which the side gate is split into four separate gates. We can observe pronounced level mixing at magnetic (B-) field induced crossings in the single particle spectrum of the constituent QDs in our device, and by applying different combinations of voltages to the four gates, we are able to perturb the effective QD confinement potentials and hence influence the level mixing. We focus on one three level crossing and one four level crossing high up in the energy spectrum of one of the QDs and describe how their energy dispersions with B-field change when different combinations of voltages are applied to the four gates. 

\section{Device Structure and Measurement Principle}
\label{structure}
\begin{figure}
\begin{center}
\includegraphics[width=8cm]{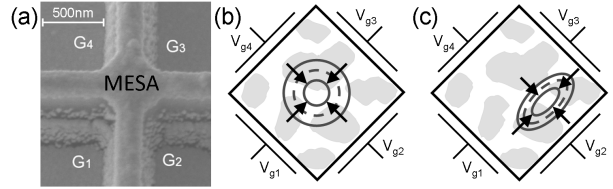}
\caption{(a) SEM image of a device mesa similar to the measured one with the location of the four gates marked as G$_{1-4}$. (b) Plan view cartoon showing an approximately circular QD (realized by applying equal voltages to all four gates) and how it is squeezed (in the direction of the arrows) by changing the gate voltages. (c) Cartoon illustrating how by application of different voltages to the four gates one can change the shape and position of the QD. The way the QD is squeezed is also different from that in (b). The feint gray irregular shapes in (b) and (c) represent the anharmonicity and anisotropy present in the device which influence the effective lateral confinement potential. Note that although our device consists of two vertically coupled QDs, for clarity, only the top-most QD is illustrated in each of the cartoons.}
\label{gates} 
\end{center}
\end{figure}

A scanning electron microscope (SEM) image of a device mesa similar to the one used in this study is shown in Fig. \ref{gates}(a). The weakly coupled vertical QD molecule is located inside the sub-micron mesa (square-shaped and at the center of the image) fabricated from a GaAs/Al$_{0.22}$Ga$_{0.78}$As/In$_{0.05}$Ga$_{0.95}$As triple barrier structure \cite{AustingPhysB1998}. Electrons are strongly confined in the vertical direction by the heterostructure barriers. The weaker lateral confinement is provided by depletion from the mesa side wall and can be changed by applying voltages to the four side gates around the mesa. The thin line mesas radiating out from the center mesa pattern the gate metal and facilitate metal contacts to bonding pads \cite{AustingSemi1997}. To study the level mixing we capture the single particle energy spectrum of one of the constituent QDs in the device by performing magneto-resonant-tunneling spectroscopy with a B-field applied parallel to the direction of current flow (out-of-dot-plane) \cite{OnoPhysB2002, PayettePRL2009, AmahaPSS2008}. This is done by using the 1s-like state in the upstream QD to probe the energy spectrum of the downstream QD in the single electron tunneling regime \cite{OnoPhysB2002}. We note that in our device the tunnel coupling energy is sufficiently small ($<$0.1meV) that we can effectively picture the states in one dot to be uncoupled from the states in the other dot, i.e., we are studying left dot and right dot states rather than bonding and anti-bonding states. 

In order to understand the influence of the four gates on the QDs, consider the cartoons in Figs. \ref{gates}(b) and (c). For simplicity, we will assume that all four gates are identical and that the action of each gate on the QDs is the same. Additionally, we will initially assume the QD confinement potentials to be ideal. In Fig. \ref{gates}(b), we first imagine that all four gates are set to the same voltage ($V_{g1}=V_{g2}=V_{g3}=V_{g4}$), giving approximately circular QDs whose centers are at the center of the mesa. Then, as the gate voltages are all simultaneously made more negative the QDs are squeezed as shown. Next, in Fig. \ref{gates}(c), suppose that the various gate voltages are initially set differently (for example, in the figure $V_{g1}\sim V_{g4}<V_{g2}\sim V_{g3}$). This can influence the QD confinement potentials in two ways. Firstly, the center of the QD confinement potentials is shifted away from the center of the mesa. Secondly, the shape of the QD confinement potentials can be changed to be approximately elliptical. After the gate voltages are initially set, they can also be all swept simultaneously to more negative voltages, which squeezes the QDs as shown. Crucially, however, the preceding discussion neglects the influence of anharmonicity and anisotropy, due to local imperfections and randomness in the structure, on the confinement potentials of realistic QDs. Previously, we have shown how these perturbations, which are represented by the gray irregular shapes in Fig. \ref{gates}(b) and (c), induce pronounced energy level mixing in the single particle spectra of our QDs \cite{PayettePRL2009, AmahaPSS2008, PayetteAPL2009}. Although the presence of these perturbations makes it hard to realize QD potentials which are nearly ideally circular or elliptical and parabolic, we we can still expect that by altering how the gate voltages are initially applied and subsequently swept, the electrons traversing the QDs will see different effective confinement potential \emph{landscapes} which can influence the coherent level mixing that occurs in the vicinity of the crossing points in the spectra. 

\section{Discussion}
\begin{figure}[t]
\begin{center}
\includegraphics[width=8cm]{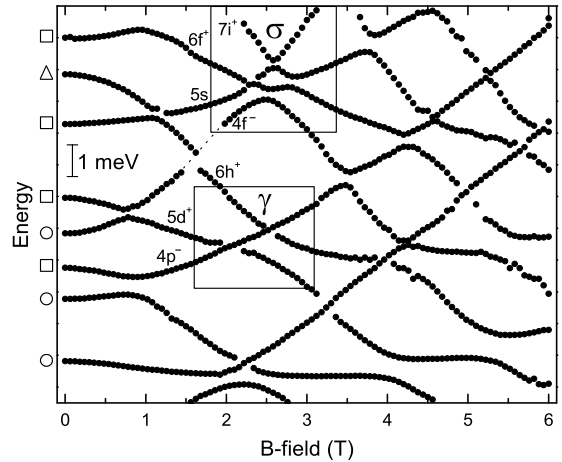}
\caption{Single particle energy spectrum of one dot from the four-gate device captured with V$_{g1}$=V$_{g2}$=V$_{g3}$=V$_{g4}$ at 1.5K. A dotted line indicates the position of a portion of one barely resolvable line in the spectrum. States relevant to the three and four level crossings of interest are labeled, for simplicity, using familiar atomic-like notation for ideal circular parabolic confinement. The symbols near the left axis help identify the nature of the states at 0T. States marked by the same symbol would be degenerate at 0T and belong to the 4th, 5th and 6th (circle, square and triangle) shells if the confining potential were exactly circular and parabolic \cite{TaruchaPRL1996, ReimannRMP2002}.} 
\label{spectra}
\end{center} 
\end{figure}

The single particle energy spectrum of one of the constituent QDs in the four-gate device, measured with equal voltages on all four gates, is shown in Fig. \ref{spectra}. One might expect that this spectrum is the closest approximation to the spectrum which would have been measured had the device only had one side gate surrounding the mesa. The energy axis in Fig. \ref{spectra} corresponds to a covariation of the bias voltage applied across the coupled QDs and the gate voltages, i.e., both higher bias and further squeezing of the QDs are required to probe levels at higher energy. Excluding the regions where two or more energy levels approach each other, the measured spectrum can be well reproduced by an energy spectrum for ideal elliptical and parabolic (two-dimensional) confinement \cite{Fock1928, Darwin1931, MadhavPRB1994, ReimannRMP2002} with confinement strengths $\hbar\omega_x=7.5$ meV and $\hbar\omega_y=5.5$ meV, i.e., the ellipticity $\delta=\omega_x/\omega_y\sim1.4$. In the vicinity of the level crossings pronounced anticrossing behavior is prevalent, and this along with the strong variation of the resonance currents (not shown in Fig. \ref{spectra}) are clear signatures of coherent mixing induced by perturbations in the probed QD's confinement potential. We note that in our earlier work on the single-gate device we were able to reproduce satisfactorily the observed anticrossing behavior at specific level crossings by including appropriate higher degree terms (from symmetry arguments) as perturbations to the probed QD confinement potential \cite{PayettePRL2009,PayetteAPL2009}.

In order to understand the energy level dispersion at three level crossings exhibiting anticrossing behavior, we have developed a simple matrix Hamiltonian model as described in Ref. \cite {PayettePRL2009}. In this model, the couplings between each pair of approaching (basis) levels, induced by the confinement potential perturbations, are characterized by the off-diagonal matrix elements. The model predicts that four basic types of energy level dispersions are possible at a three level crossing dependent on the number of dominant couplings \cite{PayettePRL2009}. In our previous experimental work, these coupling parameters could not be altered in-situ. With the new device we are now able to alter to a certain degree the coupling parameters at a given crossing by altering how the four gate voltages are swept in order to probe states arising from different QD potential \emph{landscapes}. To demonstrate this we will consider two different crossings in the spectrum of the same QD sampled differently. 

\begin{figure}[t]
\begin{center}
\includegraphics[width=8cm]{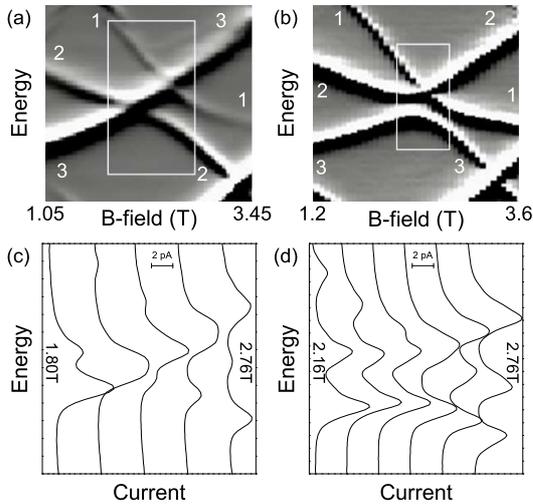}
\caption{Three level crossing $\gamma$ altered by application of different combinations of gate voltages. Panels (a) and (b) show the measured energy level (differential conductance resonance) position for two different QD potential \emph{landscapes} (both of which differ from the \emph{landscape} sampled to capture the spectrum in Fig. \ref{spectra}). Black, gray and white correspond to positive, zero and negative differential conductance. For ease of discussion, the resonances are labeled 1-3. Panels (c) and (d) show some of the current traces which come from the boxed regions in (a) and (b) [non-resonant background current not subtracted].}
\label{3level}
\end{center}
\end{figure}

We focus first on the three level crossing between the 4p$^-$, 5d$^+$ and 6h$^+$ -like states labeled $\gamma$ in Fig. \ref{spectra}. Figs. \ref{3level}(c) and (d) show explicitly some of the current traces we measured at this crossing for two different QD potential \emph{landscapes}. However, it is much clearer to see the behavior at the crossing region by taking the derivative of the current with respect to energy (effectively the differential conductance) and studying the resulting grayscale plots [Figs. \ref{3level}(a) and (b)] in which the peaks in the current traces appear as black-and-white striped lines. In Fig. \ref{3level}(a) we see three distinct resonances (labeled 1-3). As the B-field is increased resonances 1 and 2 clearly anticross while resonance 3 appears to cross the other two resonances exactly. Detailed data analysis (not shown) reveals that there are in fact small, barely resolvable anticrossings when resonance 3 intersects both resonances 1 and 2 (our spectral resolution is $\sim$50$\mu$eV). Nonetheless, this crossing is still fundamentally a one dominant coupling type crossing \cite{PayettePRL2009}. In Fig. \ref{3level}(b), we see quite different behavior for the same crossing, but with a different combination of applied voltages to the gates. Now resonances 1 and 2 appear to exactly cross while resonances 3 remains a separate branch. Counter-intuitively, this behavior can arise when all three couplings are dominant \cite{PayettePRL2009}. In fact, if all the coupling parameters are equal in magnitude, the crossing between resonance 1 and 2 would be exact. This ideal situation is highly unlikely in reality, and likely there is a small, unresolvable anticrossing between resonances 1 and 2. Detailed modeling of both the energy dependence and the currents at these crossings will be given elsewhere \cite{AmahaUNP2009}.

\begin{figure}[t]
\begin{center}
\includegraphics[width=8.5cm]{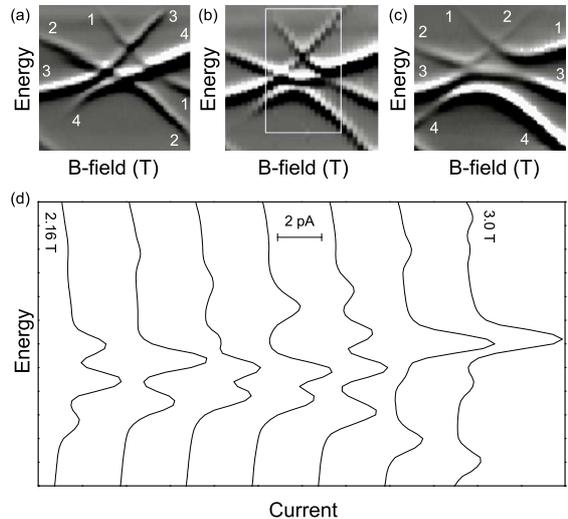}
\caption{Four level crossing $\sigma$ altered by application of different combinations of gate voltages. Panels (a), (b) and (c) show the measured energy level (differential conductance resonance) position for three different QD potential \emph{landscapes} sampled. Panel (b) is from the spectrum shown in Fig. \ref{spectra} while panels (a) and (c) are for two different QD potential \emph{landscapes}. Black, gray and white correspond to positive, zero and negative differential conductance. For ease of discussion, the resonances are labeled 1-4 in panels (a) and (c). Panel (d) shows some of the current traces from the boxed region in (b) [non-resonant background current not subtracted].}
\label{4level}
\end{center}
\end{figure}

Figure \ref{spectra} also identifies a four level anticrossing (labeled $\sigma$) between the 4f$^-$, 5s, 6f$^+$ and 7i$^+$ -like states. We are able to systematically alter the energy dispersion at this crossing, as shown in Figs. \ref{4level}(a)-(c), by applying different combination of voltages to the four gates in the device in order to sample different QD potential \emph{landscapes}. Figure \ref{4level}(a) shows an anticrossing between resonances 1 and 2 and another between resonances 3 and 4. The four ``points'' where resonances appear to cross exactly are all likely to be unresolvable anticrossings. In Fig. \ref{4level}(c) we see two distinct branches (resonances 3 and 4) and what appears to be an exact crossing between resonances 1 and 2. Fig. \ref{4level}(b) shows a case intermediate to (a) and (c) [see also selected traces in Fig. \ref{4level}(d)]. Elsewhere, we will describe how we extended our simple model for three level crossings from Ref. \cite{PayettePRL2009} to characterize four level crossings \cite{AmahaUNP2009}.

\section{Conclusion}
We have described a vertical double QD device with four side gates which allows us to alter the effective lateral confinement potential of the constituent QDs and so influence the observed level mixing in the QD energy spectra. The consequences of altering how the gate voltages are applied are clearly seen at level crossings between three and four single-particle levels. These effects, including how the resonance currents explicitly vary with B-field for different gate voltage configurations, will be more fully discussed and modeled elsewhere \cite{AmahaUNP2009}.

\section{Acknowledgments}
The authors thank B. Partoens, T. Kubo, Y. Kitamura, S. Teraoka, Y. Tokura and M. Hilke for help and discussions. Part of this work is supported by NSERC (Discovery Grant 208201), Grants-in-Aid for Scientific Research S (No. 19104007), B (No. 18340081), and by Special Coordination Funds for Promoting Science and Technology. S. T. acknowledges support from QuEST program (BAA-0824).


\end{document}